\documentclass[aps,prd,nofootinbib,showkeys,preprint,floatfix]{revtex4-1}

\usepackage{amsmath}
\usepackage{amssymb}
\usepackage{graphicx}
\usepackage{rotating}
\usepackage{ulem}
\usepackage{color} 
\usepackage{multirow} 
\usepackage{dsfont}
\usepackage[T1]{fontenc}

\newcommand{\id}{\mathds{1}}

\newcommand{\nn}{\nonumber}

\def\dblone{\hbox{$1\hskip -1.2pt\vrule depth 0pt height 1.6ex width 0.7pt
\vrule depth 0pt height 0.3pt width 0.12em$}}
\newcommand{\etmiss}{{E_T \hspace{-4.5mm}/} \hspace{2.5mm}}
\newcommand{\AddrAHEP}{
  {\it AHEP Group, Instituto de F\'{\i}sica Corpuscular --
    C.S.I.C./Universitat de Val{\`e}ncia \\
    Edificio de Institutos de Paterna, Apartado 22085,
  E--46071 Val{\`e}ncia, Spain}}

\newcommand{\AddrBonn}{%
Bethe Center for Theoretical Physics \& Physikalisches Institut der 
Universit\"at Bonn, \\
Nu{\ss}allee 12, 53115 Bonn, Germany }

\newcommand{\AddrWur}{%
Institut f\"ur Theoretische Physik und Astronomie,
Universit\"at W\"urzburg\\
Am Hubland,
97074 Wuerzburg}

\def\gsim{\raise0.3ex\hbox{$\;>$\kern-0.75em\raise-1.1ex\hbox{$\sim\;$}}}
\def\lsim{\raise0.3ex\hbox{$\;<$\kern-0.75em\raise-1.1ex\hbox{$\sim\;$}}}

\newcommand{\eq}[1]{eq.~(\ref{#1})}
\newcommand{\eqs}[2]{eqs.~(\ref{#1}) and (\ref{#2})}
\newcommand{\fig}[1]{fig.~\ref{#1}}

\newcommand{\tab}[1]{Tab.~\ref{#1}}

\begin{document}

\title{Supersymmetric type-III seesaw: \\ lepton flavour
violation and LHC phenomenology}

\author{M.\ Hirsch} \email{mahirsch@ific.uv.es}\affiliation{\AddrAHEP}

\author{W.\ Porod} \email{porod@physik.uni-wuerzburg.de}\affiliation{\AddrWur}

\author{F.\ Staub}\email{fnstaub@th.physik.uni-bonn.de}\affiliation{\AddrBonn}

\author{Ch.\ Wei{\ss}} \email{Christof.Weiss@physik.uni-wuerzburg.de}\affiliation{\AddrWur}



\preprint{Bonn-Th-2012-23, IFIC/12-65}

\begin{abstract}
We study a supersymmetric version of the seesaw mechanism type-III
considering two variants of the model: a minimal version for
explaining neutrino data with only two copies of ${\bf 24}$
superfields and a model with three generations of ${\bf 24}$-plets.
The latter predicts in general rates for $\mu\to e\gamma$
inconsistent with experimental data. However, this bound can be evaded
if certain special conditions within the neutrino sector are
fulfilled. In case of two ${\bf 24}$-plets lepton flavour violation
constraints can be satisfied much easier. After specifying the
corresponding regions in the mSugra parameter space we show that under
favorable conditions one can test the corresponding flavour structures
in the leptonic sector at the LHC. For this we perform Monte Carlo
studies  for the signals taking also into account the SUSY background. 
We find that it is only of minor importance for  the scenarios studied here.

\end{abstract}

\maketitle

\section{Introduction}

Neutrino oscillation experiments give currently the main indication
for physics beyond the Standard Model (SM). The observed tiny neutrino
masses can be easily explained by the seesaw mechanism, which at tree
level can be written in just three different variants
\cite{Ma:1998dn}, classified according to the $SU(2)\times U(1)$
representation of the postulated heavy particles: type I postulates
fermionic gauge singlets
\cite{Minkowski:1977sc,Yanagida:1979as,GellMann:1980vs,Mohapatra:1979ia},
type II scalar $SU(2)$-triplets with hypercharge 1 
\cite{Schechter:1980gr,Cheng:1980qt} and type III fermionic triplets
in the adjoint representation of $SU(2)$ \cite{Foot:1988aq}.
At low energies they all lead to a unique dimension-5 operator 
\cite{Weinberg:1979sa,Weinberg:1980bf}
\begin{equation}\label{eq:dim5}
(m^{\nu})_{\alpha\beta} = \frac{f_{\alpha\beta}}{\Lambda} (H L) (H L) .
\end{equation}
Neutrino experiments determine only $f_{\alpha\beta}/\Lambda$, but
contain no information about the origin of this operator, nor about
the absolute size of $\Lambda$. If $f$ is a coefficient ${\cal O}(1)$,
current neutrino data indicates $\Lambda \lsim {\cal O}(10^{15})$ GeV.
This value is close to, but slightly below the scale of hypothetical great unified 
theory (GUT) which should be larger than roughly $10^{16}$ GeV to avoid bounds
from the non-observation of proton decay.

A possible way to stabilise the large hierarchy between the GUT scale
and the electroweak scale is supersymmetry \cite{Witten:1981nf}.  In
its minimal form it leads to an unification of the gauge couplings in
contrast to the SM \cite{Dimopoulos:1981yj,Ibanez:1981yh,Marciano:1981un,Einhorn:1981sx,Amaldi:1991cn,Langacker:1991an,Ellis:1990wk}.
Moreover it can explain electroweak symmetry breaking as a radiative
effect \cite{Ibanez:1982fr}. Supersymmetric variants of the different
seesaw models have been considered by several authors, see
e.g.~\cite{Hisano:1998wn,Rossi:2002zb,Buckley:2006nv,Hirsch:2008gh,%
  Borzumati:2009hu,Esteves:2010ff}.  In these models
renormalisation group evolution induce non-zero flavour mixing
elements in the mass parameters of the sleptons even if they are
flavour diagonal at the GUT scale.  These in turn leads to sizeable
contributions to lepton flavour violating (LFV) observables
\cite{Borzumati:1986qx}.  In case of seesaw type-I, low energy LFV
decays such as $l_i \to l_j +\gamma$ and $l_i \to 3 l_j$ have been
calculated in 
\cite{Hisano:1995nq,Hisano:1995cp,Ellis:2002fe,Deppisch:2002vz,Petcov:2003zb,
  Arganda:2005ji,Petcov:2005yh,Antusch:2006vw,Deppisch:2004fa,Hirsch:2008dy};
$\mu-e$ conversion in nuclei has been studied in
\cite{Arganda:2007jw,Deppisch:2005zm}. 

To maintain gauge coupling unification the seesaw particles need to be
included in complete $SU(5)$ representations, i.e.\ one needs a {\bf
  15}-plet in case of type II and at least two {\bf 24}-plets in case
of type III models.  If one were to use only one
{\bf 24}-plet, then one would need either non-renormalizable
  operators at the GUT-scale  \cite{Biggio:2010me} or an extended 
$SU(5)$ Higgs sector \cite{Perez:2007iw} to explain neutrino data. 
  The type-II and type-III models
have received less attention than type-I. Note, however, that the
former has actually fewer free parameters than type-I implying that
ratios of LFV decays of leptons can actually be predicted as a
function of neutrino angles in minimal supergravity (mSUGRA) as discussed in
\cite{Rossi:2002zb,Hirsch:2008gh}.  For type-III it has been shown in
ref.~\cite{Esteves:2010ff} that a generic model with three {\bf
  24}-plets is heavily constrained by the bounds on rare lepton
decays, in particular due to the stringent bound on $\mu\to e \gamma$.
The impact on $\mu\to e \gamma$ in a gauge mediated supersymmetry breaking (GMSB) embedding of type-III
 was studied in ref.~\cite{Mohapatra:2008wx}, while in ref.~\cite{Abada:2011mg} 
also possible LHC phenomenology from lepton-flavor violation has been discussed 
for the mSUGRA case.

In this paper we are first going to show under which conditions the 
type-III model is consistent with the experimental data. This will then
be compared with a two generation model where the bounds due to $\mu\to e
\gamma$ are less stringent. Finally we will address the question to
which extent LHC might observe lepton flavour violating processes in
supersymmetric (SUSY) cascade decays. 
 Compared to the previous study in ref.~\cite{Abada:2011mg} we do not only consider
the case of two as well as of three generations of {\bf 24}-plets,
 but  we also take into account the recently measured
reactor angle $\theta_{13}$. Moreover, we demonstrate in a Monte Carlo study for the LHC signal that
the SUSY background is well under control.

For the particle content we will assume the MSSM as framework. The
recent observation of a Higgs like state at the LHC with a mass around
125 GeV \cite{ATLAS:2012ae,Chatrchyan:2012tx} can hardly be explained
within GUT models with universal boundary conditions
\cite{Bechtle:2012zk,Buchmueller:2012hv}. The same holds in variants
including seesaw states at high scales \cite{Hirsch:2012ti}.  However,
it is well-known that singlet extension of the MSSM can more easily
explain a Higgs mass of this size as there are additional F-term
contributions already at tree-level, see
e.g.~\cite{Ellwanger:2012ke,Gunion:2012zd,Ross:2011xv} 
and references therein.  Including an additional singlet does not lead
to any significant changes in the slepton sector as the singlet Yukawa
couplings enter only at two-loop level in the RGEs of the
corresponding parameters. For this reason we do consider bounds from
direct searches at the LHC but do not take into account the
requirement of correctly explaining a Higgs mass of 125 GeV.
To explain also this condition and taking into account the theoretical
as well experimental uncertainty it would be sufficient to shift the 
tree-level Higgs mass for all points in the following by about 5~GeV. In the 
next-to minimal supersymmetric standard model (NMSSM)
the tree-level mass of the light Higgs is given by \cite{Ellwanger:2009dp}.
\begin{equation}
m^2_Z\left(\cos^2 2\beta + \frac{\lambda^2}{g^2} sin^2 2\beta\right)
\end{equation}
where we introduced the superpotential coupling $\lambda$
of the singlet to the Higgs doublets.
Thus, assuming $\tan\beta=10$ one would need $\lambda \simeq 0.6$, 
i.e. not too close to the perturbativity bound of 0.75. 
However, the exact value of $\tan\beta$ plays only a sub-dominant role for
our analysis in the following. If we choose $\tan\beta=5$, already
$\lambda=0.45$ would be sufficient. Since the singlino couples only very
weakly to the sleptons its role is negligible as long as it isn't the
LSP. However, it can only be the LSP if the trilinear self coupling $\kappa$ of the singlet field
is much smaller than $\lambda$ \cite{Choi:2004zx}. For example in the scale invariant NMSSM a 
bino LSP is a common feature \cite{Cheung:2012qy}.
Moreover, this constraint can easily be satisfied in more general singlet extensions
with an explicit bilinear singlet term in the superpotential \cite{Delgado:2012yd,Ross:2012nr}.

This paper is organised as follows: in the next section we summarise
the main features of the two variants of the type-III model. In section 
\ref{sec:numerics} we first discuss how to accommodate the rare
lepton decays in type III seesaw models. Afterwards we discuss
lepton flavour violating signals from SUSY cascade decays and present
the results of a Monte Carlo study. Finally we draw in section
\ref{sec:conclusion} our conclusions.

\section{Models and spectra}
\label{sec:model}

In this section we briefly summarise the main features of the
supersymmetric version of the seesaw type-III model. In order to
maintain gauge coupling unification for type-III we add at the seesaw
scale(s) additional particles to obtain complete $SU(5)$
representation, i.e.\ a 24-plet. Note that the 24-plet actually also
includes a gauge singlet and, thus, one has always a combination of
the type I and the type III seesaw in this model.

In the subsequent sections we present the superpotentials and the
relation of the parameters to neutrino physics. In addition, there are
the corresponding soft SUSY breaking terms which, however, reduce at the
electro-weak scale to the MSSM ones and, thus, are not discussed
further. There are additional terms of the soft SUSY breaking
potential, due to the heavy particles, that we do not discuss either,
as their effect is at most of the order $M_{EWSB}/M_{seesaw}$ and,
thus, can be safely neglected. 

In this paper we will assume common soft SUSY breaking parameters at the
GUT-scale $M_{GUT}$ to specify the spectrum at the electro-weak scale:
a common universal gaugino mass $M_{1/2}$, a common scalar mass $m_0$
and the trilinear coupling $A_0$ which gets multiplied by the
corresponding Yukawa couplings to obtain the trilinear couplings in
the soft SUSY breaking Lagrangian. In addition the sign of the $\mu$
parameter is fixed, as is $\tan\beta =v_u/v_d$ (at the electro-weak
scale), where $v_d$ and $v_u$ are the the vacuum expectation values
(vevs) of the neutral component of $H_d$ and $H_u$, respectively. The
models discussed below also contain new bilinear parameters in the
superpotential leading to additional bilinear terms in the soft SUSY
breaking potential which are proportional to $B_0$ of the MSSM Higgs
sector. The corresponding RGEs decouple and their only effect is a
small mass splitting between the new heavy scalar particles from the
new heavy fermionic states of the order $B_0/M_{seesaw}$. This leads
to a tiny effect in the calculation of the thresholds at the seesaw
scale(s) \cite{Kang:2010zd} which, however, we can safely neglect.

\subsection{Supersymmetric seesaw type-III}
\label{sec:modelIII}

In the case of a seesaw model type-III one needs new fermions $\Sigma$ 
at the high
scale belonging to the adjoint representation of $SU(2)$. This has to be 
embedded in a {\bf 24}-plet  to obtain a complete $SU(5)$  representation.
The superpotential of the unbroken $SU(5)$ relevant for our discussion is
\begin{eqnarray}\label{eq:spot5}
W & = & \sqrt{2} \, {\bar 5}_M Y^5 10_M {\bar 5}_H 
          - \frac{1}{4} 10_M Y^{10} 10_M 5_H 
  +  5_H 24_M Y^{III}_N{\bar 5}_M +\frac{1}{2} 24_M M_{24}24_M \thickspace.
\end{eqnarray}
Here we have not specified the Higgs sector responsible for the
$SU(5)$ breaking as it only enters logarithmically via
threshold corrections at the GUT-scale and, thus,
plays a minor r\^ole for the subsequent discussion.
The new parts, which will give the seesaw mechanism, come from 
the $24_M$. It decomposes under  $SU(3)\times SU(2) \times U(1)$ as 
\begin{eqnarray}\label{eq:def24}
24_M & = &(1,1,0) + (8,1,0) + (1,3,0) + (3,2,-5/6) + (3^*,2,5/6) \thickspace, \\ \nn
   & = & \widehat{B}_M + \widehat{G}_M + \widehat{W}_M + \widehat{X}_M + \widehat{\bar X}_M \thickspace.
\end{eqnarray}
The fermionic components of $(1,1,0)$ and $(1,3,0)$ have exactly 
the same quantum numbers as a right-handed neutrino $\nu^c$ and 
the required $SU(2)$-triplet $\Sigma$. Thus, the $24_M$ 
always produces a combination of the type-I and type-III seesaws. 

In the $SU(5)$ broken phase the superpotential becomes
\begin{eqnarray}\label{eq:spotIII}
 W_{III} & = &  W_{MSSM}
 +  \widehat{H}_u( \widehat{W}_M Y_N - \sqrt{\frac{3}{10}} 
               \widehat{B}_M Y_B) \widehat{L}
 + \widehat{H}_u \widehat{\bar X}_M Y_X \widehat{D}^c \nonumber \\
         & & + \frac{1}{2} \widehat{B}_M M_{B} \widehat{B}_M 
         + \frac{1}{2}\widehat{G}_M M_{G} \widehat{G}_M 
          + \frac{1}{2} \widehat{W}_M M_{W} \widehat{W}_M 
          + \widehat{X}_M M_{X} \widehat{\bar X}_M \, .s
\end{eqnarray}
As before we use at the GUT scale the boundary condition
$Y_N = Y_B = Y_X$ and $M_B = M_G=M_W=M_X$. $Y_N$, $Y_B$ and $Y_X$ are
$n \times 3$ while $M_G$, $M_W$ and $M_X$ are $n \times n$-dimensional matrices if 
we include $n$ generations of 24-plets. Integrating out the heavy fields
yields the following formula for the neutrino masses at the low scale:
\begin{equation}
m_\nu = - \frac{v^2_u}{2} 
\left( \frac{3}{10} Y^T_B M^{-1}_B Y_B + \frac{1}{2} Y^T_W M^{-1}_W Y_W \right). 
\label{eq:mnu_seesawIII}
\end{equation}
As mentioned above there are two contributions, one from the gauge
singlet the other from the $SU(2)$ triplet.  In this case the
calculation of the Yukawa couplings in terms of a given high scale
spectrum is more complicated than in the other two types of seesaw
models. However, as we start from universal couplings and masses at
$M_{GUT}$ we find that at the seesaw scale one still has $M_B \simeq
M_W$ and $Y_B \simeq Y_W$ so that one can write in a good
approximation
\begin{equation}
m_\nu = - v^2_u  \frac{4}{10} Y^T_W M^{-1}_W Y_W \,.
\label{eq:mnu_seesawIIIa}
\end{equation}

Being complex symmetric, the light Majorana neutrino mass matrix
in \eq{eq:mnu_seesawIIIa}, is diagonalized by a unitary 
$3\times 3$ matrix  $U$~\cite{Schechter:1980gr}
\begin{equation}\label{diagmeff}
{\hat m_{\nu}} = U^T \cdot m_{\nu} \cdot U\ .
\end{equation}

Inverting the seesaw equation, \eq{eq:mnu_seesawIIIa}, 
allows to express 
$Y_W$ as  \cite{Casas:2001sr}
\begin{equation}\label{Ynu}
Y_W =\frac{i 4 \sqrt{2}}{5 v_u}\sqrt{\widehat{W}_M}\cdot R \cdot \sqrt{{\hat
    m_{\nu}}} \cdot U^{\dagger},
\end{equation}
for $n=3$ where the $\hat m_{\nu}$ and $\widehat{W}_M$ are diagonal
matrices containing the corresponding eigenvalues.  $R$ is in general
a complex orthogonal matrix which is characterised by three angles
$\phi_i$ which are in general complex. Note that, in the special case
$R={\bf 1}$, $Y_{W}$ contains only ``diagonal'' products
$\sqrt{M_im_{i}}$. For $U$ we will use the standard form
\begin{eqnarray}\label{def:unu}
U=
\left(
\begin{array}{ccc}
 c_{12}c_{13} & s_{12}c_{13}  & s_{13}e^{-i\delta}  \\
-s_{12}c_{23}-c_{12}s_{23}s_{13}e^{i\delta}  & 
c_{12}c_{23}-s_{12}s_{23}s_{13}e^{i\delta}  & s_{23}c_{13}  \\
s_{12}s_{23}-c_{12}c_{23}s_{13}e^{i\delta}  & 
-c_{12}s_{23}-s_{12}c_{23}s_{13}e^{i\delta}  & c_{23}c_{13}  
\end{array}
\right) 
 \times
 \left(
 \begin{array}{ccc}
 e^{i\alpha_1/2} & 0 & 0 \\
 0 & e^{i\alpha_2/2}  & 0 \\
 0 & 0 & 1
 \end{array}
 \right)
\end{eqnarray}with $c_{ij} = \cos (\theta_{ij})$ and $s_{ij} = \sin (\theta_{ij})$. The angles $\theta_{12}$,
 $\theta_{13}$ and  $\theta_{23}$ are the solar neutrino angle, the reactor (or CHOOZ) angle
and the atmospheric neutrino mixing angle, respectively. $\delta$ is the Dirac phase
and $\alpha_i$ are Majorana phases. In the following we will set the latter to 0 and
consider for $\delta$ mainly the cases $0$ and $\pi$.

\subsection{Supersymmetric seesaw type-III with two \textbf{24}-plets}
\label{224}

Current neutrino experiments only determine the differences of the
neutrino masses squared. Thus it might well be that only two of the
light neutrinos are massive whereas the third is either massless or
has a mass much smaller than the others. Such a situation is obtained
if only two \textbf{24}-plets are present similarly to the case of the
seesaw type I with two right-handed neutrinos as discussed for example
in \cite{ibarra2m24,2m24b,2m24c}.  We call this class of models
$3\times 2$ seesaw, see also \cite{masiero_vives}.

In the following we work in the basis where  $M_W$ is a $2\times 2$ 
diagonal matrix denoting the eigenvalues by $\hat M_i$ ($i=1,2$).
Similarly to the three generation case one can express the
$Y_W$ in terms of low-energy neutrino parameter and model dependent
high energy parameters as also discussed in the context
of seesaw I models \cite{Ibarra:2005qi}:
\begin{equation}
				Y_{W}=\sqrt{\frac{5}{2}}\frac{i}{v_u}\sqrt{M_W}R\left(\sqrt{\widehat{m}_{\nu}^{-1}}\right)'U^{\dagger}.
			\label{YukIII2}
\end{equation}	
The $R$-matrix is now $2\times 3$ which can assume the following forms:
for normal hierarchy ($m_1=0$)
\begin{equation}
 R_{\mathrm{norm}}= \begin{pmatrix} 0 & \cos(\phi) & -\sin(\phi) \\ 0 & \sin(\phi) & \cos(\phi) \end{pmatrix} 
\label{R2normal}
\end{equation}
in case of normal hierarchy in the neutrino sector ($m_1=0$) or
and for inverse hierarchy ($m_3=0$)
\begin{equation}
 R_{\mathrm{norm}}= \begin{pmatrix} \cos(\phi) & -\sin(\phi) & 0\\  \sin(\phi) & \cos(\phi) & 0\end{pmatrix} 
\label{R2inverse}
\end{equation}
in case of  inverse hierarchy ($m_3=0$). Note that the $R$-matrix
is parametrised by one complex angle $\phi$ only in contrast to
the three generation case.

\subsection{Lepton flavour violation in the slepton sector}
\label{sec:model_LFV}

From a one-step integration of the RGEs one gets, assuming mSUGRA
boundary conditions, a first rough estimate for the lepton flavour
violating entries in the slepton mass parameters:
\begin{eqnarray}
\label{eq:LFVentriesML}
({\Delta}m^2_{L})_{ij} &\simeq& -\frac{a_k}{8 \pi^2 } 
 \left( 3 m^2_0 +  A^2_0 \right) 
 \left(Y^{k,\dagger}_W L Y^{k}_W\right)_{ij} \thickspace, \\
({\Delta}A)_{l,ij} &\simeq& -a_k \frac{3}{ 16 \pi^2 }   A_0 
 \left(Y_e Y^{k,\dagger}_W L Y^{k}_W\right)_{ij} \thickspace,
\label{eq:LFVentriesA}
\end{eqnarray}
for $i\ne j$ in the basis where $Y_e$ is diagonal, 
$L_{ij} = \ln(M_{GUT}/M_i)\delta_{ij}$ and $Y^k_W$ is the additional Yukawa 
coupling where $k$ indicates the number of \textbf{24}-plets. 
\begin{equation}
a_2= \frac{6}{5} \text{ and } a_3 = \frac{9}{5} \, .
\label{eq:LFVentriesII}
\end{equation}
Both models have in common that they predict negligible flavour
violation for the right-sleptons
\begin{eqnarray}
({\Delta}m^2_{E})_{ij} &\simeq& 0
\end{eqnarray}
which is a general feature of the usual seesaw models 
\cite{Esteves:2010ff}. Although it is known that approximations
\eqs{eq:LFVentriesML}{eq:LFVentriesA} do not
reproduce well the actual size of the off-diagonal elements
they do give the functional dependencies on the high scale parameters.
 Therefore they are a useful
indicator on how the rare lepton decays $l_i \to l_j \gamma$ depends
on these parameters as the corresponding decay modes
scale  roughly like
\begin{equation}
Br(l_i \to l_j \gamma) \propto \alpha^3 m_{l_i}^5
	\frac{| ({\Delta}m^2_{L})_{ij}|^2}{\widetilde{m}^8}\tan^2\beta.
\label{eq:LLGapprox}	
\end{equation}
where $\widetilde{m}$ is the average of the SUSY masses involved in the 
loops. 
Using the parametrization for the Yukawa couplings of eq.~(\ref{Ynu})
the entries in $(\Delta m_L^2)_{ij}$ can be expressed as
\begin{equation}\label{wq:delmgen}
(\Delta {m_L^2})_{ij} 
\propto U_{i\alpha}U_{j\beta}^*\sqrt{m_{\alpha}}\sqrt{m_{\beta}}
R_{k\alpha}^*R_{k\beta}M_k\log\left(\frac{M_X}{M_k}\right).
\end{equation}

In the special case that the matrix $R$ is the identity matrix, 
eq.~(\ref{wq:delmgen}) reduces to
\begin{eqnarray}
\label{delmR1}
\left(\Delta{m_{{L}}^2}\right)_{12} & 
\propto & c_{12} c_{13} \left( - s_{12}c_{23} - c_{12}s_{23}s_{13} 
e^{-i\delta} \right) z_1 \\ \nonumber
& + & s_{12}c_{13} \left(c_{12}c_{23} - s_{12}s_{23}s_{13} 
e^{-i\delta} \right) z_2 + s_{23}c_{13}s_{13} e^{-i\delta}  z_3  \\ \nonumber
\left(\Delta{m_{{L}}^2}\right)_{13} & 
\propto & c_{12} c_{13} \left(s_{12} s_{23} - c_{12}c_{23}s_{13} 
e^{-i\delta} \right) z_1 \\ \nonumber
& + & s_{12}c_{13} \left( - c_{12}s_{23} - s_{12} c_{23}s_{13} 
e^{-i\delta} \right) z_2 + c_{23}c_{13}s_{13} e^{-i\delta} z_3  \\ \nonumber
\left(\Delta{m_{{L}}^2}\right)_{23} & 
\propto  & \left( s_{12}s_{23} - c_{12}c_{23}s_{13} 
e^{-i\delta} \right) \left(-s_{12}c_{23} - c_{12}s_{23}s_{13} 
e^{i\delta} \right) z_1 \\ \nonumber
& + & \left( - c_{12}s_{23} - s_{12}c_{23} s_{13} 
e^{-i\delta} \right) \left( c_{12}c_{23} - s_{12}s_{23}s_{13} 
e^{i\delta} \right) z_2 \\ \nonumber
& + & s_{23}c_{23}c_{13}^2 z_3
\end{eqnarray}
where 
\begin{equation}
z_i\equiv m_i M_i \log\left(\frac{M_X}{M_i}\right) \, .
\end{equation}
For the ansatz of degenerate seesaw states the combination
$M_i\log (\frac{M_X}{M_i})$ becomes an overall factor, i.e. 
for degenerate $M_B=M_W$ one may simply make
the replacement $z_i \to m_{i}$ in eq.~(\ref{delmR1}).
For strict normal hierarchy, the expressions become even simpler. For instance,
$\Delta{m_{{L}}^2}$ becomes
\begin{eqnarray}
\left(\Delta{m_{{L}}^2}\right)_{12} & 
\propto \left( s_{13}s_{23} \sqrt{\Delta(m^2_{\rm Atm})+\Delta(m^2_{\odot})}
-\sqrt{\Delta(m^2_{\odot})}s_{12}^2+c_{12}c_{23}e^{i\delta}
\sqrt{\Delta(m^2_{\odot})}s_{12}
\right),
\end{eqnarray}
Inserting the best fit point data for oscillation 
parameters, except for $s_{13}$, and assuming $\delta=\pi$ one 
can calculate the value for $s_{13}^2$ for which 
$\Delta{m_{{L}}^2}$ approximately vanishes 
as $s_{13}^2=0.0077$, which agrees very well with the full 
numerical calculation shown in the next section, see Fig.~\ref{th13scans}.

Similar analytical estimates can be calculated in other limits 
and, even though absolute values for LFV processes are only 
rough estimates, ratios of LVF quantities can be calculated quite 
accurately in this way.

In the numerical studies we will use the complete formulas as given
in \cite{Bartl:2003ju,Arganda:2005ji}.
 We will also consider the three body decays 
BR$(l_i \to 3 l_j)$ where we use the formulas given in
\cite{Arganda:2005ji}.

\section{Numerical results}
\label{sec:numerics}

In this section we present our numerical calculations. All results
presented below have been obtained with the lepton flavour violating
version of the program package SPheno \cite{Porod:2003um,Porod:2011nf}. The
RGEs of the two seesaw III models have been calculated with
\texttt{SARAH} \cite{Staub:2008uz,Staub:2009bi,Staub:2010jh,Staub:2012pb}. 
For the Monte Carlo studies below we have used the \texttt{SUSY Toolbox}
\cite{Staub:2011dp} to generate the interface to \texttt{WHIZARD}
\cite{Kilian:2007gr}.

All seesaw parameters as well as the soft SUSY breaking parameters
are defined at  $M_{GUT}$.
  We evolve the RGEs to the scales corresponding
to the GUT scale values of the masses of the heavy particles. The RGE
evolution implies also a splitting of the heavy masses  up to
20\% between the gauge singlet and the color octet. We therefore
add at the corresponding scale the threshold effects due to the heavy
particles to account for the different masses as discussed in 
\cite{Esteves:2010ff}. However, since the gauge singlet doesn't 
contribute to the running of the gauge couplings, the main impact on
gauge coupling unification is due to mass splitting between the color 
octet and the $SU(2)_L$ triplet. This splitting is for a seesaw scale of 
$O(10^{14}~\text{GeV})$ of the order of 10\% and would result in a 
marginal shift of $O(10^{-4})$ for the gauge couplings. 
Off-diagonal elements 
are induced in the mass matrices of the ${\bf 24}$-plets. This
implies that one has to go the corresponding mass eigenbasis before
calculating the threshold effects.  We use two-loop RGEs everywhere
except stated otherwise.  

Unless mentioned otherwise, we fit neutrino mass squared differences
to their best fit values \cite{Tortola:2012te}. 
Our numerical procedure is as follows: inverting 
the seesaw equation, see \eqs{Ynu}{YukIII2}, one can get a first guess for the Yukawa
couplings for any fixed values of the light neutrino masses (and
angles) as a function of the corresponding triplet mass for any fixed
value of the couplings.  This first guess will not give the correct
Yukawa couplings, since the neutrino masses and mixing angles are
measured at low energy, whereas for the calculation of $m_\nu$ we need
to insert the parameters at the high energy scale. However, it can be used
 to run numerically the RGEs to obtain the exact
neutrino masses and angles (at low energies) for these input
parameters. The difference between the results obtained numerically
and the input numbers can then be minimized in a simple iterative
procedure until convergence is achieved.  As long as neutrino Yukawas
are $|Y_{W, ij}|< 1$ $\forall i,j$ we reach convergence in a few steps.

\subsection{Bounds from lepton decays}

It is known for some time that generically the supersymmetric seesaw
III model predicts rates for $\mu \to e \gamma$ which are too large
\cite{Esteves:2010ff} to be compatible with the experimental bound for
BR$(\mu\to e \gamma) \lsim 2.4 \cdot 10^{-12}$ \cite{Adam:2011ch}.
However, this does not completely exclude the model as there are
certain parameter regions where cancellations between different
contributions can occur. In this section we explore the different
possibilities. For the corresponding regions the question arises if
they can be probed by other experiments, in particular the LHC.  From
the discussion in the previous section, in particular
\eqs{eq:LFVentriesML}{eq:LLGapprox}, the rare
leptons decays are mainly governed by the overall SUSY mass scale and
the lepton flavour entries. The LFV entries in the softs are nearly
completely governed by the choice of parameter in the heavy seesaw
sector, while the dependence on the soft SUSY parameters is much
weaker. We therefore fix the later to:
\begin{equation}
m_0=1000~\text{ GeV}\,,\,\,
M_{1/2}=1000~\text{ GeV}\,,\,\,
A_0=0\,\text{ and }
\mu> 0 \,.
\label{eq:susypar}
\end{equation}
 
\begin{figure}[t]
  \centering
    \includegraphics[scale=1.10]{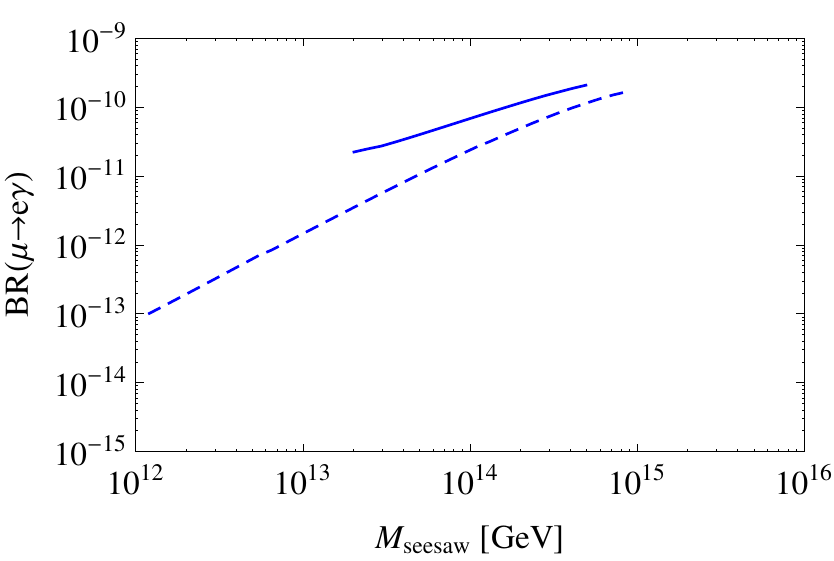}   
  \caption{BR($\mu\to e\gamma$) versus the seesaw scale for the two 
  (dashed) and the three (solid) \textbf{24}-plet scenario; mSUGRA 
  parameters like in \eq{eq:susypar}, neutrino data fixed to the current 
  experimental values, including $\sin^2\theta_{13}=0.026$.}
  \label{seesawscale23}
\end{figure}
In \fig{seesawscale23} we recall the generic situation for the
type-III seesaw. The dashed and full line correspond to the 2- and
3-generation model, respectively. Only a certain range for $M_{\rm
  Seesaw}$ is allowed. The lower bounds stem from the fact that the
gauge couplings become non-perturbative at the GUT-scale whereas the
upper bounds are due to non-perturbative Yukawa couplings at the
GUT-scale \cite{Esteves:2010ff}. Every $\bf 24$-plet contributes with
$\Delta b_i=5$ to the beta functions of the gauge couplings $g_i$ and,
thus, obviously the possible range is larger for the 2-generation case
compared to the 3-generation case.

\begin{figure}[t]
\centering
\begin{minipage}[c]{6.8cm}
\centering
\includegraphics[scale=1]{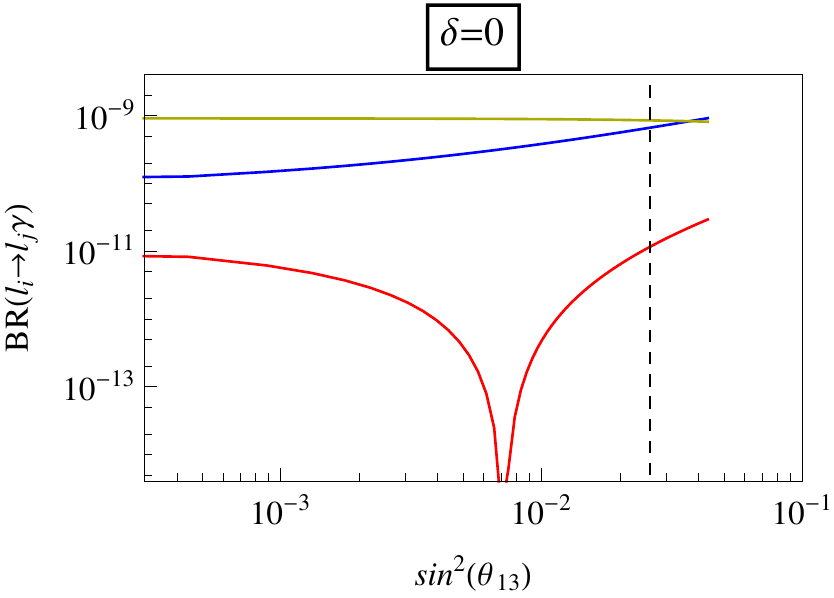}
\end{minipage}
\hspace{2.5cm}
\begin{minipage}[c]{6.8cm}
\centering
\includegraphics[scale=1]{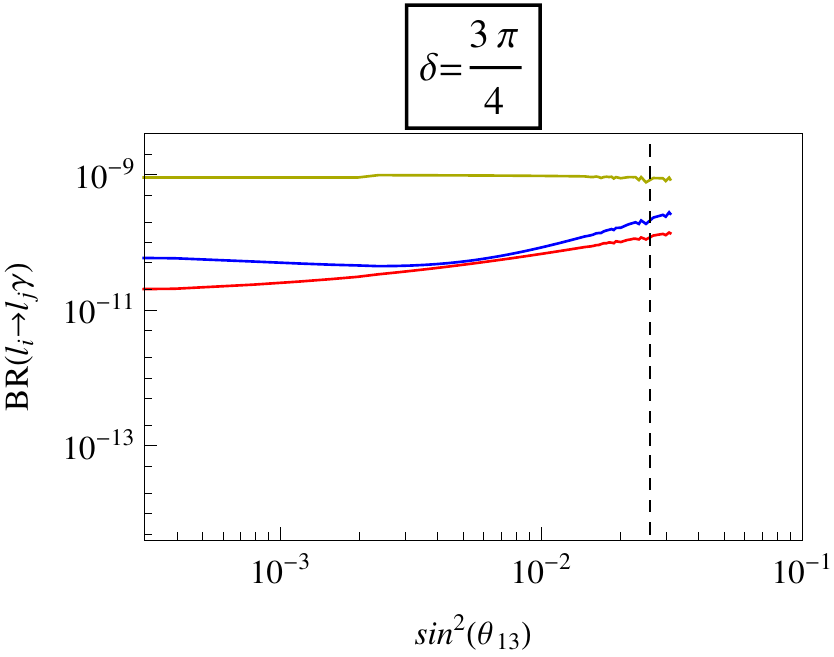}
\end{minipage}
\begin{minipage}[c]{6.8cm}
\centering
\includegraphics[scale=1]{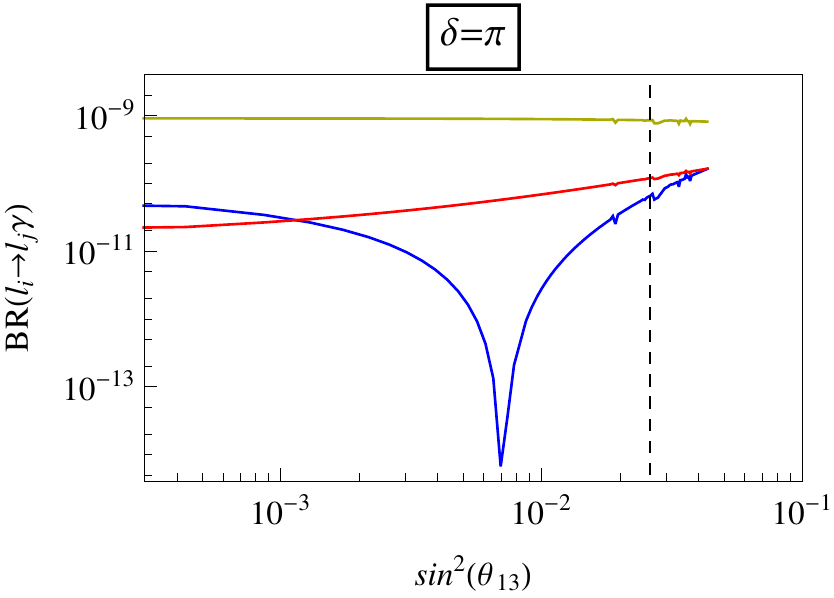}
\end{minipage}
\caption{BR($\mu\to e\gamma$) (blue), BR($\tau\to e\gamma$) (red) and 
BR($\tau\to\mu\gamma$) (yellow) over the reactor angle $\theta_{13}$ for 
real parameters with Dirac phases $\delta=0$ (upper left), $\delta=\pi$ 
(upper right) and $\delta=3\pi/4$ (lower panel). 
We set $M_W = 10^{14} \cdot \dblone_3$ and the SUSY parameters
as in \eq{eq:susypar}.
The dashed line indicates the 
current best-fit value for $\theta_{13}$.}
\label{th13scans}
\end{figure}

Equations (\ref{Ynu}) and (\ref{YukIII2}) imply that one  can induce
special features for the Yukawa couplings when varying
$\sin\theta_{13}$, the CP-phases and/or elements of the $R$-matrix as
has also been noted in ref.~\cite{Antusch:2006vw} in case of
supersymmetric seesaw I models.  This has an immediate impact on the
flavour mixing entries of the slepton mass parameters as can be seen
from \eqs{eq:LFVentriesML}{eq:LFVentriesA}. As an example we show in
\fig{th13scans} the dependence on $\theta_{13}$ in the range allowed
before the results of DAYA-BAY \cite{An:2012eh} and RENO
\cite{Ahn:2012nd} assuming three different values for the Dirac-phase
$\delta$ and a degenerate mass of $10^{14}$ GeV for the ${\bf
24}$-plets.  Note that in this particular case the elements of
$R$-matrix do not play any r\^ole. As can be seen, $\delta$ has to be
close to $\pi$ in this case to get BR$(\mu\to e \gamma)$ below the
experimental bound.  For completeness we note that the small spikes in
the plots are numerical artifacts of our iterations procedure.

\begin{figure}[t]
\centering
    \includegraphics[scale=0.9]{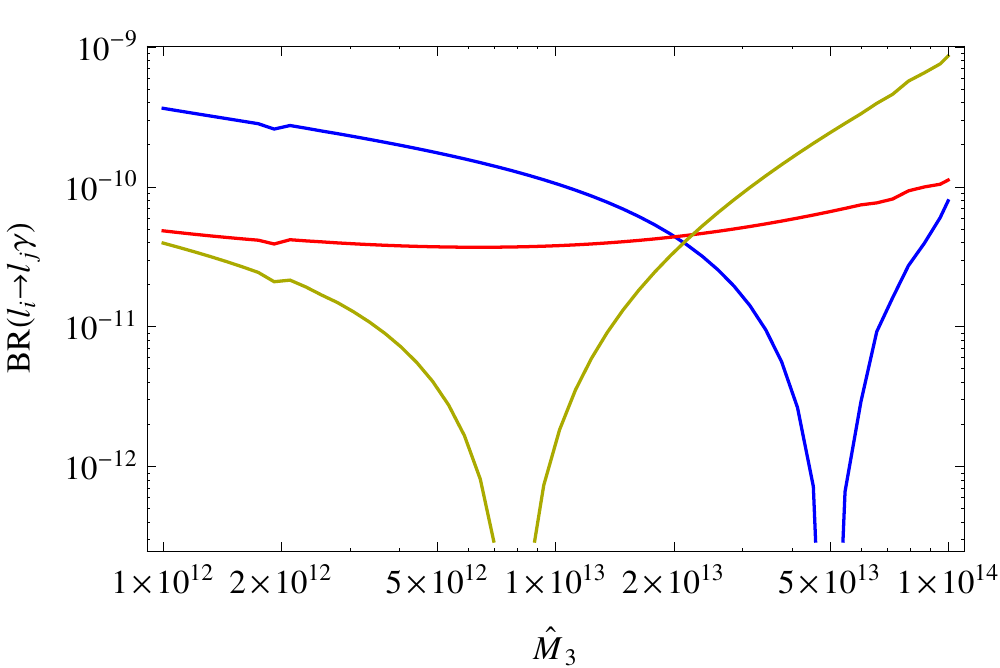}   
\caption{BR($\mu\to e\gamma$) (blue), BR($\tau\to e\gamma$) (red) and 
  BR($\tau\to\mu\gamma$) (yellow) as function of $\hat M_3$ with $\theta_{13}$ at the current best-fit value. We have taken 
$\delta=\pi$, $\hat M_1=\hat M_2=10^{14}$ GeV and the other parameters
 as in \eq{eq:susypar}.}
  \label{th13_hierarch}
\end{figure}

\begin{figure}[t]
\centering
    \includegraphics[scale=1.10,trim=0 0 0 22,clip]{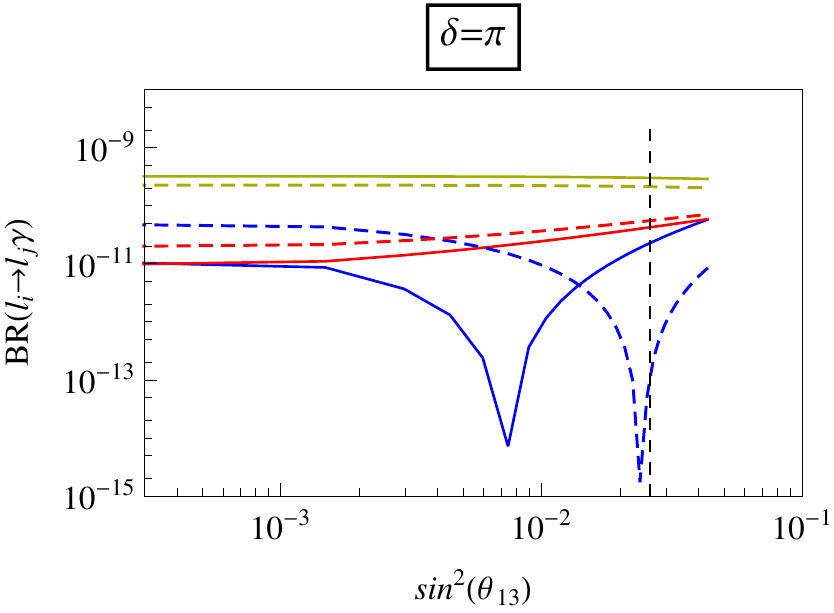}   
  \caption{BR($\mu\to e\gamma$) (blue), BR($\tau\to e\gamma$) (red) and 
  BR($\tau\to\mu\gamma$) (yellow) as  function of $\sin^2\theta_{13}$
  in the 2-generation model for $\delta=\pi$,  
  $\hat M_1=\hat M_2=10^{14}$~GeV (solid lines) and  
  $\hat M_1=2\cdot 10^{14}$~GeV, $\hat M_2=10^{14}$~GeV (dashed lines).
  The other parameters are as in \eq{eq:susypar}.}
  \label{th13_2m24}
\end{figure}

With the recent measurement of $\theta_{13}$ by Daya Bay and RENO one
can now relate seesaw parameters from the requirement to respect the
bound on $\mu\to e \gamma$.\footnote{ For a recent update on $\mu \to
  e \gamma$ in seesaw type I models taking into account the measured
  value of $\theta_{13}$ see \cite{Calibbi:2012gr}.} As an example we
fix in \fig{th13_hierarch} $\delta=\pi$, 
$\hat M_1 = \hat M_2=$
$10^{14}$ GeV and take $R=\dblone_3$. In this case the bound on $\mu
\to e \gamma$ is satisfied if $\hat M_3$ is close to $5\cdot 10^{13}$
GeV.  Note however, that the numbers obtained depend on the SUSY point
chosen in parameter space. Therefore, one can start to constrain the
seesaw parameters only after the discovery and subsequent
determination of the SUSY parameters.  In \fig{th13_2m24} we show a
similar graph but for the two generation model. The interesting point
is that despite the fewer parameters one still has sufficient freedom
to suppress the rare lepton decays.  One the one hand, this shows the
need to determine not only the differences of the neutrino masses
squared but also the absolute neutrino mass scale or in other words
the mass of the lightest neutrino, as a non-zero value of the latter
would rule out the minimal two generation model or requires its extension
by non-renormalizable operators. On the other hand it implies
that for the exploration of the LHC phenomenology it is sufficient to
study the simpler two generation model.

\begin{figure}[t]
\includegraphics[scale=0.87,trim=0 0 0 21,clip]{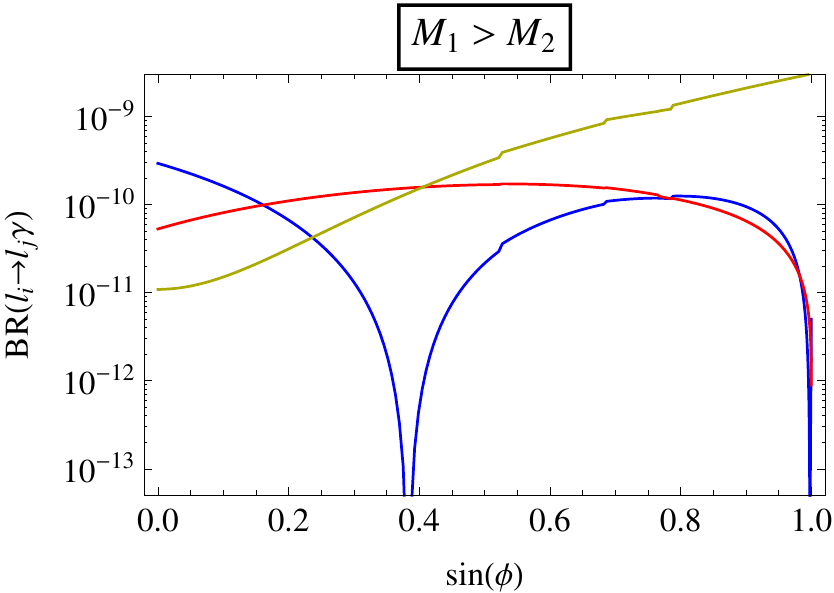}
\includegraphics[scale=0.9,trim=0 0 0 21,clip]{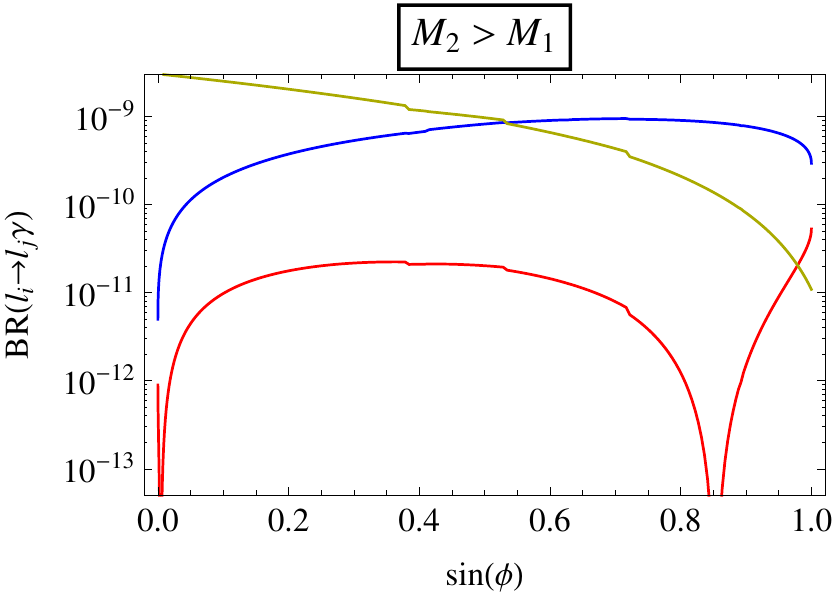}
\caption{BR($\mu\to e\gamma$) (blue), BR($\tau\to e\gamma$) (red) and 
  BR($\tau\to\mu\gamma$) (yellow) as function of $\sin \phi$
for two \textbf{24}-plets with masses $\hat M_1=5\cdot 10^{14}$~GeV and 
$\hat M_2=5\cdot 10^{13}$~GeV on the left panel (right panel vice versa $\hat M_1 > \hat M_2$), $\delta=0$ and the SUSY parameters as in
\eq{eq:susypar}.}
\label{r2m24normal}
\end{figure}

Up to now we have assumed that the $R$-matrix is the unit matrix.
In \fig{r2m24normal} we study the dependence on the $R$-matrix
in the two-generation model. We fix the \textbf{24}-plet masses to
$5\cdot 10^{13}$~GeV and $5\cdot 10^{14}$~GeV and vary $\sin(\phi)$. 
Note that in both cases we have taken $\cos \phi >0$. Instead of
taking $\hat M_1 > \hat M_2$ we could have taken $\cos \phi <0$
in the second plot. As expected, we find that variation of the
$R$-matrix provides additional possibilities to suppress BR($\mu\to e \gamma$)
below the current experimental bound. Moreover, our results show
that one never can exclude this class of models by these measurements
as with a sufficient tuning of parameters one can always avoid the
bounds. 

\begin{figure}[th]
\includegraphics[scale=0.9]{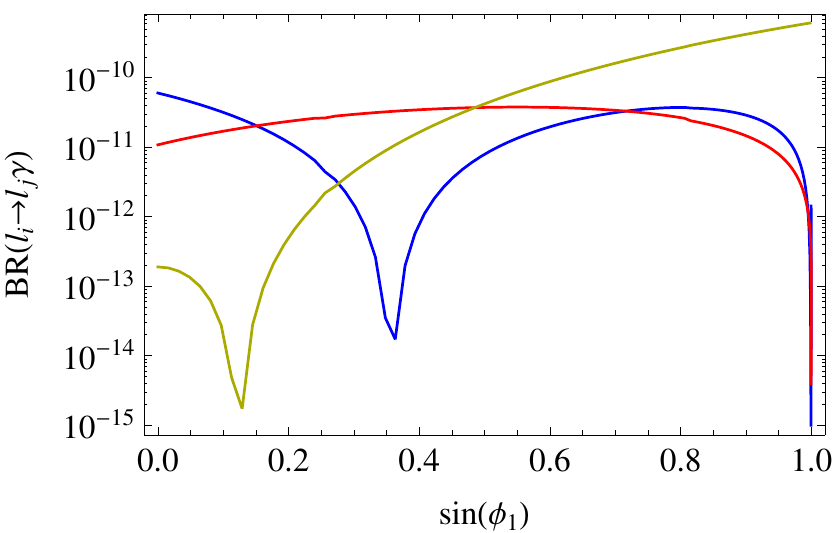}\hspace{0.2cm}
\includegraphics[scale=0.9]{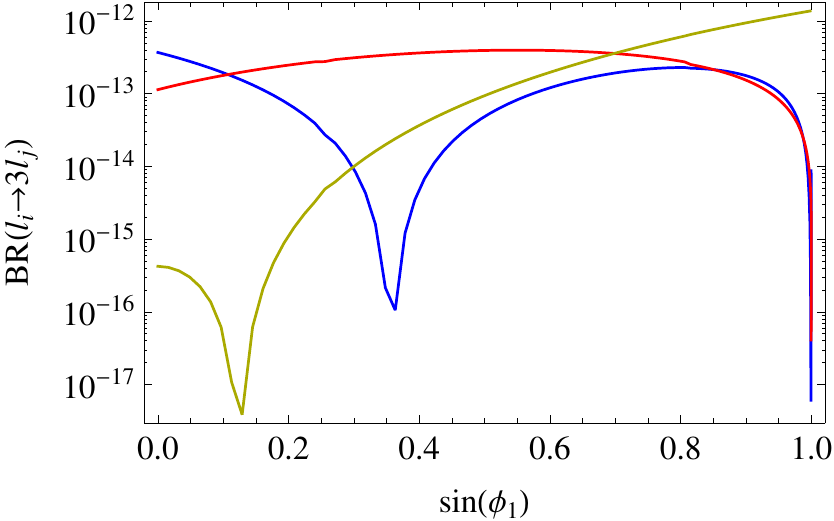}
\caption{Comparison of the two-body decays $l_i\to l_j\gamma$ (left panel) and the-three body decays $l_i\to 3 l_j$ (right panel) for variation of $\sin(\phi_1)$ at normal neutrino mass hierarchy,  Dirac phase $\delta=0$ and a \textbf{24}-plet hierarchy: $\hat M_1= 10^{15}$~GeV, $\hat M_2= 10^{14}$~GeV  and
$\hat M_3=10^{13}$~GeV; $l_i$, $l_j$ = $\mu$, $e$ (blue); $\tau$, $e$ (red); $\tau$, $\mu$ (yellow).}
\label{r3body}
\end{figure}

For completeness we compare in \fig{r3body} the branching ratios of
the two body decays $l_i \to l_j \gamma$ to the ones for the three
body decays $l_i \to 3 l_j$ in the three-generation model.  Similar to
the seesaw type I case \cite{Arganda:2005ji} we see that both decay
classes show the same dependence on the underlying parameters, since in
case of the three body decays the photon contribution dominates.  We
have checked that this also holds for the two-generation model.

\subsection{Testing flavour structures at the LHC}

We have seen in the previous section that one can choose the
seesaw parameters such that the experimental constraints on
the rare leptons decays are fulfilled. In this section we address
the question if there are any possibilities to test these models
for such parameter choices at the LHC. As we will demonstrate 
there are indeed favorable SUSY parameter regions where 
one can observe the corresponding flavour violating decays 
of supersymmetric particles.
 
The branching ratios of the lepton flavour violating decays of sleptons
and neutralinos are governed by the same entries in the slepton
mass matrix as the rare lepton decays, i.e.\ the ones given in
\eqs{eq:LFVentriesML}{eq:LFVentriesA}.
Therefore both classes of decays show the same dependence on the seesaw
parameters.

At the LHC one has to study cascade decays containing sequences
of the form 
$\tilde \chi^0_2 \to \tilde l_k^\pm l^\mp_j \to l_i^\pm l^\mp_j\tilde \chi^0_1$ 
with $i\ne j$
\cite{Hinchliffe:2000np,Deppisch:2004pc,Bartl:2005yy,Andreev:2006sd%
,delAguila:2008iz,Hirsch:2008dy,Carquin:2008gv,Esteves:2009vg}.
Moreover, the nature of the neutralinos should be dominantly
gaugino like and the mass difference should be small enough to
suppress the decay into $h^0$. This 
requires a certain hierarchy between the neutralino mass parameters
and the slepton mass parameters which is roughly given by
$|\mu| \gg M_2 \gsim m_{\tilde l} \gsim M_1$ where the ordering of
$M_1$ and $M_2$ can be interchanged.

As the scaling of the lepton flavour violating decays of SUSY
particles is similar to the one of the rare lepton decays in this
class of models we use the following strategy to enhance the rates for
$\tilde \chi^0_2 \to \tilde e^\pm \mu^\mp \chi^0_1$, $\tilde \chi^0_2
\to \tilde e^\pm \tau^\mp \chi^0_1$ and $\tilde \chi^0_2 \to \tilde
\mu^\pm \tau^\mp \chi^0_1$: for a given point in the SUSY parameter
space we choose the seesaw parameters in the following way: we fix the
$R$-matrix to be either $\dblone$ or as in \eq{R2normal}, depending
whether we work in the two- or three-generation seesaw model.  Next we
fix the relative size of various entries of $Y_W$ such that the
neutrino mixing matrix is tribimaximal. Note, that a non-zero
$\theta_{13}$ changes the neutralino branching ratios only slightly 
and, thus, its effect can be neglected her.
In the third step both $Y_W$
and $M_W$ are rescaled until the correct neutrino masses are obtained
and $10^{12}\cdot \text{BR}(\mu\to e \gamma)$ is in the interval
$[2.2,2.4]$. In this way one obtains the maximum rate for the decay
$\tilde \chi^0_2 \to \tilde e^\pm \mu^\mp \chi^0_1$ which is the
cleanest at the LHC \cite{Andreev:2006sd,delAguila:2008iz}.  With a
further variation of the entries in $Y_W$ one could increase the final
states containing a $\tau$ lepton. However, we have checked for couple
of points in the SUSY parameter space that this would only lead to an
relative increase of about ten per-cent for the corresponding
rates. We have not pursued this further as this is at most of the
order of the expected theoretical uncertainty on the SUSY cross
section, which is about 10-20 per-cent, see
e.g.\ \cite{Beenakker:2011sf,Kramer:2012bx,Hollik:2012rc,Langenfeld:2012ti}
and refs.\ therein. In the following examples we have checked that the
bounds on SUSY particles are fulfilled
\cite{Chatrchyan:2012jx,CMS:2012mfa,ATLAS:2012rz,ATLAS:2012ms}.

\begin{figure}[t]
\includegraphics[scale=0.9]{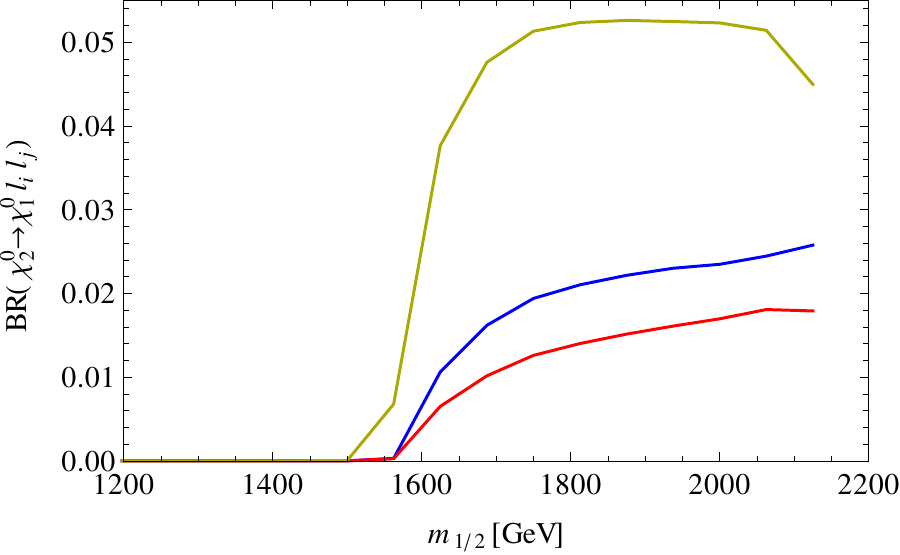}\hspace{0.2cm}
\includegraphics[scale=0.9]{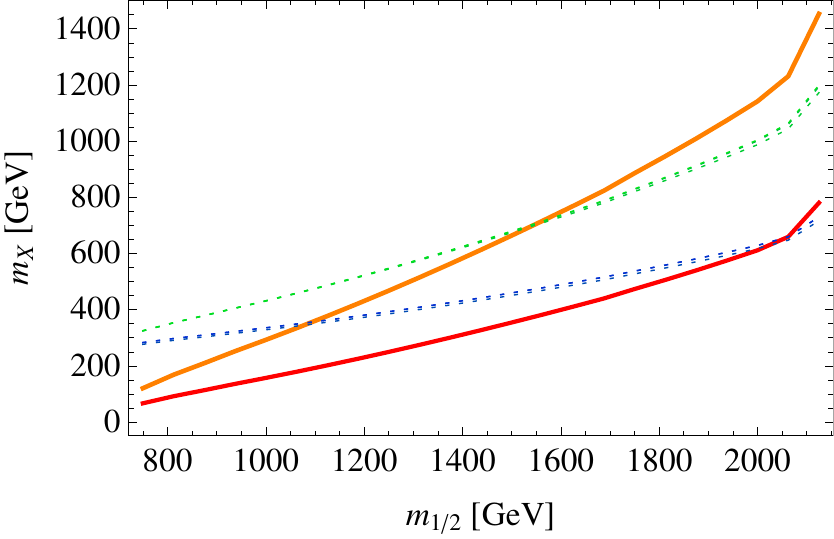}
\caption{BR($\widetilde{\chi}^0_2\to\widetilde{\chi}^0_1 l_i l_j$)
and selected masses as function of $M_{1/2}$ for $m_0=250$~GeV,
$A_0=0$, $\tan\beta=10$ and $\mu >0$. Left plot:
BR($\widetilde{\chi}^0_2\to\widetilde{\chi}^0_1 \mu e$) (blue), 
BR($\widetilde{\chi}^0_2\to\widetilde{\chi}^0_1 \tau e$) (red) and 
BR($\widetilde{\chi}^0_2\to\widetilde{\chi}^0_1 \tau\mu$) (yellow);
right plot: $\widetilde{\chi}^0_2$ (orange), $\widetilde{\chi}^0_1$ (red),
$\widetilde{l}_{1,2,3}$ (blue dotted) and $\widetilde{l}_{4,5,6}$
(green dotted).
The neutrino parameters are at tri-bi-maximal values, normal 
neutrino mass hierarchy and $R=\id$; $M_W$ is varied to fit 
BR($\mu\to e\gamma$) close to the experimental bound.}
\label{spectrum}
\end{figure}
 
\begin{figure}[t]  \centering
    \includegraphics[scale=1.10]{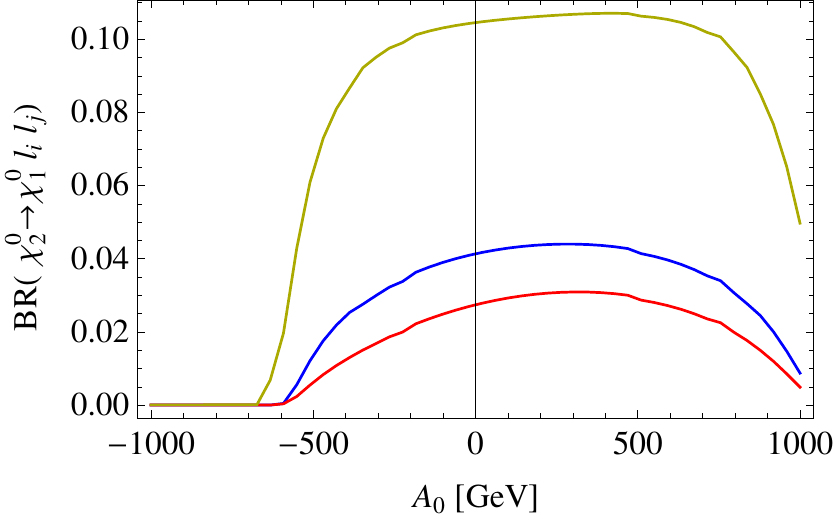}   
  \caption{BR($\widetilde{\chi}^0_2\to\widetilde{\chi}^0_1 \mu e$) (blue), 
  BR($\widetilde{\chi}^0_2\to\widetilde{\chi}^0_1 \tau e$) (red) and 
  BR($\widetilde{\chi}^0_2\to\widetilde{\chi}^0_1 \tau\mu$) (yellow) 
in the 2-generation model
as function of $A_0$ for $m_0=250$~GeV, $M_{1/2}=1800$~GeV, 
$\tan\beta=10$ and $\mu >0$. The seesaw parameters are fixed as explained
in the text.}
  \label{A0scan}
\end{figure}

The branching ratios of the lepton flavour
violating decays can reach up to a few per-cent as shown in
\fig{spectrum}. The structure of the RGEs implies that the
three heaviest sleptons are essentially $\tilde l_L$ even
though there can be sizeable mixing between the stau states.
The latter mixing is the main source of the LFV decays for 
$M_{1/2} \lsim 1550$~GeV where only the three lightest sleptons
appear in the $\tilde \chi^0_2$ decays. The hierarchy
BR($\widetilde{\chi}^0_2\to\widetilde{\chi}^0_1 \tau\mu) >$ 
BR($\widetilde{\chi}^0_2\to\widetilde{\chi}^0_1 \mu e ) >$ 
BR($\widetilde{\chi}^0_2\to\widetilde{\chi}^0_1 \tau e$) is a
consequence of the structure of $Y_W$ needed to explain the neutrino
data. The change of the spectrum has two main sources: (i) 
$M_{1/2}$ enters the RGEs for the slepton mass parameters. 
(ii) The requirement that BR$(\mu\to e \gamma)$ to be in above
interval implies that the seesaw scale becomes a function of
$M_{1/2}$. Changing the seesaw scale has a major impact on
spectrum as discussed in detail in ref.~\cite{Esteves:2010ff}.
Similar features show up in the 2-generation model as
exemplified in \fig{A0scan} where we show the LFV $\widetilde{\chi}^0_2$
decay branching ratios as a function of $A_0$. In this model
one can find LFV branching ratios of up to 10 per-cent. The main
reason for this are the different kinematics for the same
mSUGRA input because changing the number of seesaw particles 
implies changes in the RGEs  of the slepton and gaugino mass
parameters as discussed above.

\begin{figure}[t]
\centering
\includegraphics[scale=0.85]{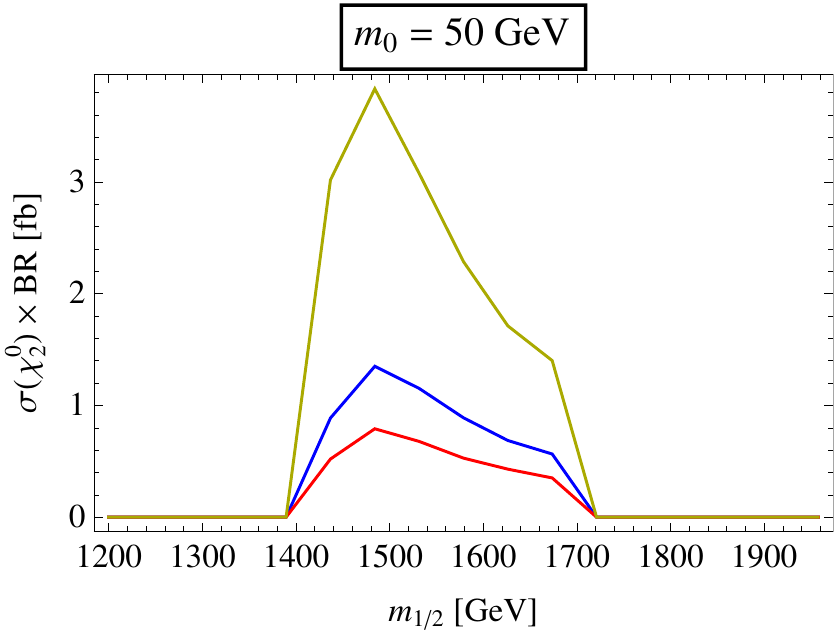}   
    \includegraphics[scale=0.85]{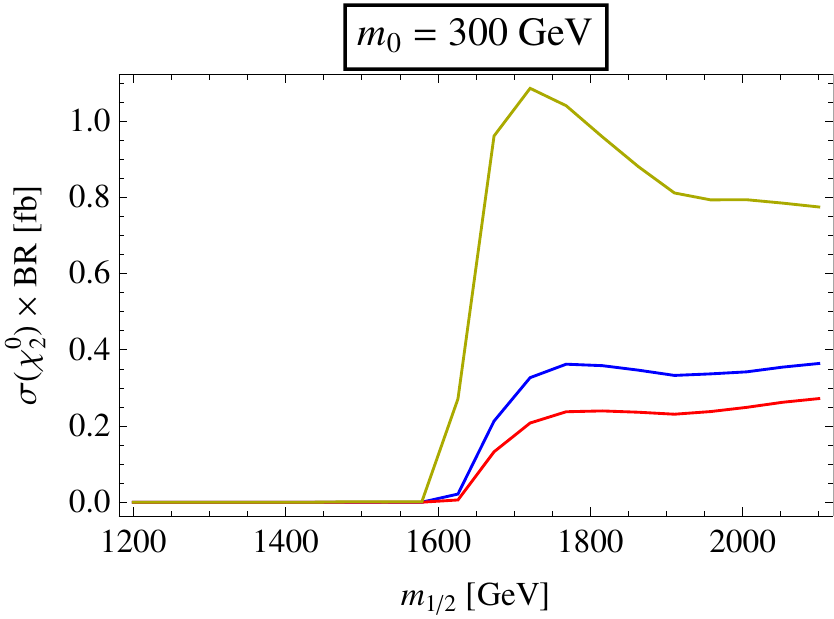}
  \caption{Grid calculation of $\sigma(\widetilde{\chi}^0_2)\times\text{BR}
(\widetilde{\chi}^0_2\to\widetilde{\chi}^0_1 \mu e$) (blue), $\sigma(\widetilde{\chi}^0_2)\times\text{BR}
(\widetilde{\chi}^0_2\to\widetilde{\chi}^0_1 \tau e$) (red) and $\sigma(\widetilde{\chi}^0_2)\times\text{BR}
(\widetilde{\chi}^0_2\to\widetilde{\chi}^0_1 \tau\mu$) (yellow) in femtobarn over $M_{1/2}$ for different 
values of $m_0$; $A_0=0$, 
$\tan\beta=10$ and $\mu >0$. The seesaw parameters are fixed as explained
in the text.}
  \label{grid}
\end{figure}

We concentrate in the following on the two generation model
as here the signal is somewhat larger than in the three generation
model.
At the LHC $\widetilde{\chi}^0_2$ is mainly produced in
the cascade decays of squarks and gluinos. In \fig{grid}
we show $\sigma\times$BR as a function of $M_{1/2}$ fixing
the other  parameters for two values of $m_0$, $A_0=0$,
$\tan\beta=10$, $\mu>0$ and $\sqrt{s}=14$~TeV. In addition,
 we used 
$\hat{M}_1 = \hat{M}_2 = 2.5\cdot 10^{13}$~GeV as well as 
$Y_{W,11} = Y_{W,12}=-Y_{W,13}=-5.252\cdot 10^{-2}$, $Y_{W,21}=0$  and
$Y_{W,22}=Y_{W,23}=-1.547\cdot10^{-1}$.
Here we have summed
over all possibilities to produce squarks and gluinos and we
require that the two leptons from  $\widetilde{\chi}^0_2$ are
the only ones in the event. For the calculation of the cross section
we have used the \texttt{LHC-FASER} package \cite{Dreiner:2010gv,benfaser}. 
One sees that
the signal cross section before putting any cuts can be at most
a few fb which gives at most a few hundred events even for an
integrated luminosity of 300 fb$^{-1}$. 

\begin{figure}[t]
\centering
\includegraphics[scale=0.8]{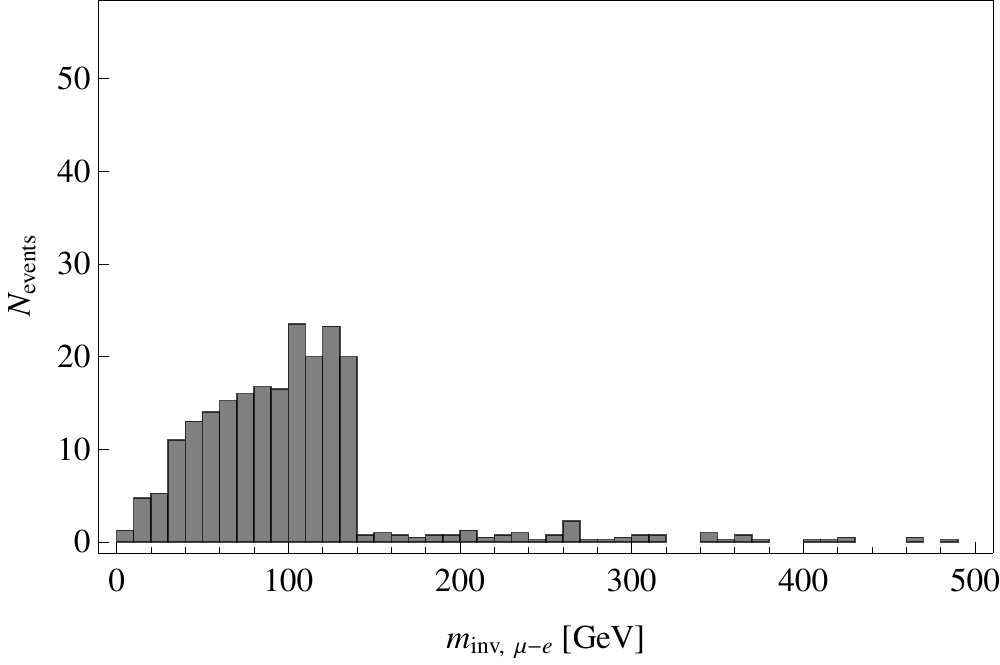}   
\includegraphics[scale=0.8]{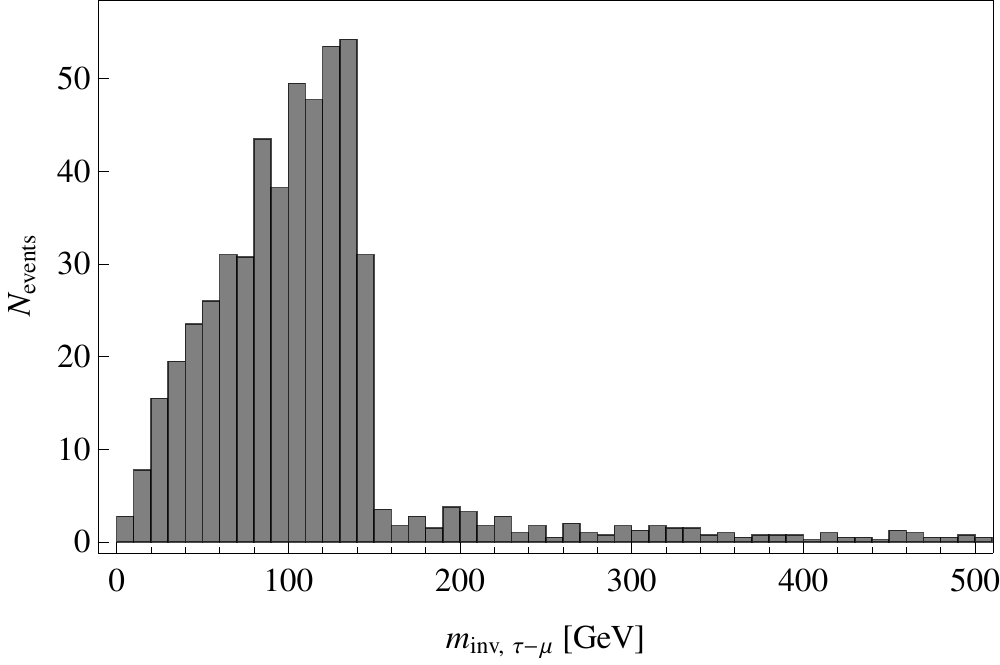}   
  \caption{Invariant mass distributions for the signal and 
 SUSY background for final states containing
  $\mu$, $e$ (left plot) or $\tau$, $\mu$ (right plot),
  missing transverse energy and at least two jets in the final state for 
  a luminosity of $300\text{ fb}^{-1}$, $m_0=50$~GeV, $M_{1/2}=1484$~GeV,
$A_0=0$, $\tan\beta=10$  and $\mu>0$.}
  \label{wh_mue}
\end{figure}

This naturally leads to the question if such a signal can be
observed at all. For this reason we have performed a Monte Carlo
study at the parton level taking $m_0=50$~GeV, $M_{1/2}=1484$~GeV,
$A_0=0$, $\tan\beta=10$  and $\mu>0$ corresponding to maximum 
of the signal in \fig{grid}. For the generation of the events we
use \texttt{WHIZARD} \cite{Kilian:2007gr}.
The corresponding signal cross sections
are 1.4, 0.8 and 3.8 fb for the final states containing $e\mu$, $e\tau$ and 
$\mu\tau$, respectively. 
Related  studies have been performed
in refs.~\cite{Hinchliffe:2000np,Hisano:2002iy,Andreev:2006sd} where
it has been shown that one can reduce the SM background sufficiently.
However, to our knowledge the SUSY background has not yet been taken
into account. This will be also considered here.
The main SM background are due $t\bar{t}$, $VV$
and $Vjj$ ($V=W,Z$)  production. 
In ref.~\cite{juliaharz}, where a detector
study for the $\mu\tau$ channel has been performed
for $\sqrt{s}=10$~TeV, it has been shown
that the SM background can reduced significantly by requiring
a cut on the missing transverse energy $\etmiss >140$~GeV
and a cut on the effective mass $M_{eff} > 400$~GeV 
where
\begin{equation*}
M_{eff}\equiv  \etmiss + \sum_{i=1}^4 p^{jet}_T \, .
+\sum_j p^{l}_T
\end{equation*}
The first sum is over the transverse momentum of the four hardest
jets and the second one over the transverse momentum of all leptons.
We have adjusted these cuts for the
case  $\sqrt{s}=14$~TeV and use $\etmiss >150$ $M_{eff} > 1200$~GeV.
Moreover we require that the event contains exactly two leptons
and no $b$-jets. 
This reduces the SM-background to a negligible level. The main SUSY
background is due to charginos and $W$-bosons produced in the SUSY 
cascade decays. In contrast to the signal these events stem in
general from cascade decays of different squarks and/or gluinos.
Therefore, if one plots the differential cross section as
a function of the invariant lepton mass $m_{ll'}=\sqrt{(p_l+p_{l'})^2}$
one gets a triangle for the peak and a flat distribution
from the background. We have simulated the combination of signal with
SUSY background using the dominant production mechanism which is in
this case squark-squark production as the squarks are much lighter than
the gluino yielding about 80 per-cent of the total cross section.
The results for an integrated luminosity of 300 fb$^{-1}$ are shown
in \fig{wh_mue} where we have cut the range of $m_{ll'}$ at 500 GeV
even though the SUSY background continues flat until about 1 TeV.
As can be seen, one gets approximately the triangular shape of
the signal with the edge at
\begin{equation}
m^2_{ll'} = \frac{(m^2_{\tilde \chi^0_2} - m^2_{\tilde l})
                  (m^2_{\tilde l}-m^2_{\tilde \chi^0_2})}
                 {m^2_{\tilde l}}
\label{eq:egde} 
\end{equation}
where the lepton masses have been neglected. The edge clearly
indicates the consecutive two  body decays giving a first hint
on the mass ordering.
As the sleptons
have different masses, they give somewhat different values for the
edges which are collected in \tab{tab:edges}.
\begin{table}
\caption{Edges, as given in \eq{eq:egde}, of the invariant lepton masses
due to the individual sleptons. The two neutralino masses are
344.6~GeV and 647.0~GeV}
\label{tab:edges}
\begin{tabular}{|c|c|c|}\hline
slepton & mass [GeV] & $m_{ll'}$ [GeV] \\ \hline
$\tilde l_1$ & 377.9 & 213.0 \\
$\tilde l_{2,3}$ & 386.0 & 233.9 \\
$\tilde l_4$ & 621.9 & 148.3 \\
$\tilde l_5$ & 625.1 & 139.3 \\
$\tilde l_6$ & 625.9 & 136.7 \\ \hline 
\end{tabular}
\end{table}
Figure \ref{wh_mue} clearly shows that in this case the SUSY background
is negligible compared to the signal. Note that the light sleptons
hardly contribute to the signal as argued above and, thus, the edges
are essentially due to the heavier sleptons. In case of the $\tau\mu$
final state the two edges could be guessed but it will require
high luminosity and a finer binning to disentangle the resulting
double edge structure due to the contributions of the
different sleptons \cite{Bartl:2005yy}.

\section{Conclusions}
\label{sec:conclusion}

We have studied supersymmetric variants of the seesaw type III model.
At the electroweak scale the particle content is the same as in
the MSSM. At the seesaw scale(s) the particles have been included
in a {\bf 24}-plet to ensure unification of the gauge couplings.
In this way one ends up with a combination of seesaw type III
combined with type I where the latter gives sub-dominant contribution
if $SU(5)$-GUT conditions for the corresponding Yukawa couplings
are assumed. We have considered two variants of this model
using either two or three generation of {\bf 24}-plets. The latter
case is heavily constrained by the experimental bound on
$\mu\to e \gamma$. However, as we have shown there are various
ways to obtain cancellations between different contributions
so the bound can be respected: here the Dirac phase of the neutrino
sector enters as well as the mass hierarchy of the seesaw particles
and their mixing properties. Even though the measurement
of the reactor angle $\theta_{13}$ gives an additional constraint,
the model still has sufficient many parameters to be consistent
with all experimental data. In the two generation model the constraints
due to the rare lepton decays are less severe and can be more easily
accommodated.

We have also investigated the question to which extent lepton
flavour violating signals can be seen at the LHC. The current
experimental bounds on SUSY particles imply that within
a unified model such as the mSugra squarks and gluinos must be in the
TeV range. As the main signal is in the cascade decays of these
particles one gets at most a few fb for the signal. The corresponding
part of the parameter space is for small $m_0$ and large $m_{1/2}$
if the seesaw parameters are chosen such that BR$(\mu \to e \gamma)$
is close to its experimental bound. One can turn this around:
if the bound BR$(\mu \to e \gamma)$ is increased by an order of
magnitude than it is rather unlikely that LHC finds LFV in SUSY decays
in this class of models.

\section*{Acknowledgements}

We thank J.C.~Rom\~ao for providing
 us with his \texttt{SPheno frontend} which
facilitated the scans over the parameter space and B.~O'Leary
for providing an updated version of the \texttt{LHC-FASER} package 
for the cross section calculations.
W.P.\ and Ch.W.\ thank the IFIC for hospitality during extended stays.
Their work has been supported in part by the DFG, project no.\ PO-1337/2-1 
and the Helmholtz Alliance ``Physics at the Terascale''.
W.P.\ has been supported by the Alexander von Humboldt foundation.
M.H.\  acknowledges support from the Spanish MICINN grants
FPA2011-22975, MULTIDARK CSD2009-00064 and
by the Generalitat Valenciana grant Prometeo/2009/091 and the
EU~Network grant UNILHC PITN-GA-2009-237920.

\bibliographystyle{h-physrev5}
\bibliography{SeesawIII.bib}

\end{document}